# Giant transition-state enhancement of quasiparticle spin-Hall effect in an exchange-spin-split superconductor detected by non-local magnon spin-transport

Kun-Rok Jeon,[*] Jae-Chun Jeon, Xilin Zhou, Andrea Migliorini,

Jiho Yoon, and Stuart S. P. Parkin[*]

*Max Planck Institute of Microstructure Physics, Weinberg 2, 06120 Halle (Saale), Germany*

[*]To whom correspondence should be addressed: jeonkunrok@gmail.com, stuart.parkin@halle-mpi.mpg.de

**Although recent experiments and theories have shown a variety of exotic transport properties of non-equilibrium quasiparticles (QPs) in superconductor (SC)-based devices with either Zeeman or exchange spin-splitting, how QP interplays with magnon spin currents remains elusive. Here, using non-local magnon spin-transport devices where a singlet SC (Nb) on top of a ferrimagnetic insulator ($Y_3Fe_5O_{12}$) serves as a magnon spin detector, we demonstrate that the conversion efficiency of magnon spin to QP charge via inverse spin-Hall effect (iSHE) in such an exchange-spin-split SC can be greatly enhanced by up to 3 orders of magnitude compared with that in the normal state, particularly when its interface superconducting gap matches the magnon spin accumulation. Through systematic measurements by varying the current density and SC thickness, we identify that superconducting coherence peaks and exchange spin-splitting of the QP density-of-states, yielding a larger spin excitation while retaining a modest QP charge-imbalance relaxation, are responsible for the giant QP iSHE. The latter exchange-field-modified QP relaxation is**

**experimentally proved by spatially resolved measurements with varying the separation of electrical contacts on the spin-split Nb.**

Over the last decade, it has been shown that the combination of superconductivity with spintronics leads to a variety of novel phenomena, which do not exist separately[1-8]. In particular, recent discovery and progress in the proximity generation and control of spin-polarized triplet Cooper pairs[1-3] at carefully engineered superconductor (SC)/ferromagnet (FM) interfaces *in equilibrium* pave the way for the development of non-dissipative spin-based logic and memory technologies.

Besides triplet Cooper pairs, *non-equilibrium* quasiparticles (QPs) in a spin-split SC[4-6] have also raised considerable interest. This is because their exotic properties resulting from the mutual coupling between different non-equilibrium imbalances of spin, charge, heat, and spin-heat can greatly enhance spintronics functionality[5]. For example, the coupling of spin and heat imbalances gives rise to long-range QP spin signals as observed in Al-based non-local spin valves[9-11] with a Zeeman spin-splitting field. In addition, a temperature gradient between a normal metal (NM) and a spin-split SC separated by a tunnel barrier induces a pure QP spin current[12] without an accompanying net charge current, analogous to the spin-dependent Seebeck tunneling[13,14]. Substituting the NM by a FM, one can achieve large (spin-dependent) thermoelectric currents[15,16] far beyond commonly found in all metallic structures.

Magnon spintronics[17-19] has been an emerging approach towards novel computing devices in which magnons, the quanta of spin waves, are used to carry, transport, and process spin information instead of conduction electrons. Especially in the low-damping ferrimagnetic insulator Yttrium-Iron-Garnet ($Y_3Fe_5O_{12}$ = YIG)[20], a magnon-carried spin current can propagate over extremely long distances (centimeters at best) and it is free from Ohmic dissipation due to the absence of electrons in motion[17-19]. Despite many recent advances[17-19] in this research field, how magnon spin current interacts with and is converted to QP spin and charge currents in a spin-split SC is yet to be investigated.

In this paper, we report three key aspects of the conversion behavior of magnon spin to QP charge via inverse spin-Hall effect (iSHE) in an exchange-spin-split SC (Nb), directly probed by non-local magnon spin-transport[18] (Fig. 1a). Firstly, the iSHE in the superconducting state of Nb becomes up to 3 orders of magnitude greater than in the normal state. Secondly, this enhancement appears only in the vicinity of the superconducting transition temperature $T_c$ when the magnon spin current has an energy comparable to the (singlet) superconducting gap $2\Delta^{SC}$ of Nb (Fig. 1a). Lastly, its characteristic dependence on a d.c. current density $J_{dc}$ and the Nb thickness $t_{Nb}$ indicates that a singularity near the gap edge and a spin-splitting field are *both* essential for the giant transition-state QP iSHE, the latter of which is experimentally confirmed by performing spatial profiling of the transition state enhancement by varying the separation distance of electrical contacts on the spin-split Nb layer.

## RESULTS AND DISCUSSION

The non-local magnon spin-transport devices (Fig. 1b) we study consist of two identical Pt electrodes and a central Nb layer on top of 200-nm-thick YIG films, which are liquid-phase epitaxially grown on a (111)-oriented single-crystalline gadolinium gallium garnet ($Gd_3Ga_5O_{12}$, = GGG) wafer (see Methods). Control devices, in which a 10-nm-thick $Al_2O_3$ spin-blocking layer is inserted between Nb and YIG in an otherwise identical structure, are also prepared for comparison (Fig. 1b). Here, we send a d.c. current $I_{dc}$ through one Pt (using leads 1 and 2 in Fig. 1b) while measuring the in-plane (IP) magnetic-field-angle $\alpha$ dependence of the non-local open-circuit voltages [$V_{nl}^{Pt}(\alpha)$, $V_{nl}^{Nb}(\alpha)$] using both the other Pt (leads 7 and 8) and the central Nb (leads 3 and 4). Note that we apply an external in-plane magnetic field $\mu_0H_{ext}$ of 5 mT, larger than the coercive field of YIG (Fig. 1c), to fully align its magnetization $M_{YIG}$ along the field direction. $\alpha$ is defined as the relative angle of $\mu_0H_{ext}$ (//$M_{YIG}$) to the long axis of two Pt electrodes which are collinear.

As schematically illustrated in Fig. 1a, the right Pt acts as a normal-metal (NM) injector of magnon spin current across the Pt/YIG interface via either electron-mediated SHE (charge-to-spin conversion)[21] or spin Seebeck effect (SSE) (heat-to-spin conversion)[22] due to the accompanying Joule heating [$\Delta T \propto (I_{dc})^2$]. The left Pt serves as a NM detector of the magnon spin current, diffusing through a YIG channel, via electron-mediated iSHE (spin-to-charge conversion) whereas in the same device, the middle Nb functions as an exchange-spin-split SC detector of the diffusive magnon current via QP-mediated iSHE below $T_c$[8].

The total voltage measured across the detector is given by $V_{nl}^{tot} = \Delta V_{nl}^{el} + \Delta V_{nl}^{th} + V_0$, where $\Delta V_{nl}^{el}$ and $\Delta V_{nl}^{th}$ are proportional to the magnon spin current and accumulation created electrically (SHE $\propto I_{dc}$)[21] and thermally [SSE $\propto (I_{dc})^2$][22], respectively. These electrical and thermal magnon currents can be separated straightforwardly by reversing the polarity of $I_{dc}$, allowing us to determine the magnitude of each component based on their distinctive angular dependences[18]; $\Delta V_{nl}^{el} = \frac{[V_{nl}^{tot}(+I_{dc}) - V_{nl}^{tot}(-I_{dc})]}{2} \propto \sin^2(\alpha)$ and $\Delta V_{nl}^{th} = \frac{[V_{nl}^{tot}(+I_{dc}) + V_{nl}^{tot}(-I_{dc})]}{2} - V_0 \propto \sin(\alpha)$. $V_0$ is an offset voltage that is independent of the magnon spin-transport.

The typical result of such a measurement using the Pt detector at 300 K is displayed in Fig. 1d-1i, for the $t_{Nb}$ = 15 nm devices with and without the $Al_2O_3$ spin-blocking layer. This evidences that both electrically (Fig. 1e and 1h) and thermally (Fig. 1f and 1i) excited magnons transport spin angular momentum over a long distance of 15 µm at room temperature, which is consistent with the original work[18]. We note that from reference devices having the Pt injector/detector only, the room-temperature magnon spin-diffusion length $l_{sd}^m$ of the YIG is estimated to be around 11 (9) µm for the electrically (thermally) driven magnons (Supplementary Section 1). The transporting spin current is absorbed by the middle Nb to a certain extent, given by the difference between the signals with versus without the $Al_2O_3$ insertion (see Supplementary Section 2 for the quantitative analysis).

Figure 2a-2d shows the temperature $T$ evolution of $\Delta V_{nl}^{el}(\alpha)$ and $\Delta V_{nl}^{th}(\alpha)$ for the $t_{Nb}$ = 15 nm devices measured by the Pt detector at a fixed $I_{dc}$ = |0.5| mA ($J_{dc}$ = |3.3| $MA/cm^2$). As summarized in Fig. 2f and 2g, $\left[\Delta V_{nl}^{el}\right]^{Pt}$ diminishes with decreasing the base temperature $T_{base}$ and it almost vanishes for $T_{base} \leq 10$ K whereas $\left[\Delta V_{nl}^{th}\right]^{Pt}$ significantly increases at low $T_{base}$. Such distinct $T_{base}$-dependences are in line with previous experiments[23,24] and theoretical considerations[25,26] that the injection mechanisms for electrical and thermal magnons across the Pt/YIG interface (parameterized by the effective spin conductance and the interface spin Seebeck coefficient, respectively) differ fundamentally. Furthermore, the energy-dependent magnon diffusion and relaxation of the YIG channel may play a role in the transport process[27,28].

We below focus on the non-local signal from the thermally generated magnons ($\Delta V_{nl}^{th}$) since it remains sufficiently large at low $T_{base}$ for allowing a reliable analysis across $T_c$. In Fig. 2g, we first plot the $T_{base}$ dependence of $\left[\Delta V_{nl}^{th}\right]^{Pt}$ without the $Al_2O_3$ layer normalized by that with the Al$_2$O$_3$ layer; $\left[\Delta V_{nl}^{th}\right]^{Pt,\ no\ Al_2O_3}/\left[\Delta V_{nl}^{th}\right]^{Pt,\ with\ Al_2O_3}$. This value reflects how much the magnon spin current is absorbed by the Nb layer. Notably, $\left[\Delta V_{nl}^{th}\right]^{Pt,\ no\ Al_2O_3}/\left[\Delta V_{nl}^{th}\right]^{Pt,\ with\ Al_2O_3}$ drops abruptly right below $T_c$ (extracted from the Nb resistance $R^{Nb}$ versus $T_{base}$ plot of Fig. 2e) and then it rises progressively as entering deep into the superconducting state, resulting in a downturn at $T_{base}/T_c \approx 0.95$ (inset of Fig. 2f). Such a non-trivial behaviour is compatible with recent theoretical predictions[29,30] and experimental reports[31,32] on *ferromagnetic insulator* (FMI)/SC structures, where (spin-singlet) Cooper pairs from the SC cannot leak into the FMI even if the exchange spin-splitting can still penetrate the SC[4]. So rather well-developed coherence peaks of the QP density-of-states (DOS) at the FMI/SC interface[5] are accessible to the transporting spin current. This gives rise to an anomalous enhancement of spin absorption by the adjacent SC near $T_c$. Note that in contrast, for *metallic/conducting* FM/SC proximity-coupled structures[33], $2\Delta^{SC}$ is significantly suppressed at the FM/SC interface and the superconducting coherence peak effect is therefore fading away[29,34-38]. A slight rise in $\left[\Delta V_{nl}^{th}\right]^{Pt,\ no\ Al_2O_3}/\left[\Delta V_{nl}^{th}\right]^{Pt,\ with\ Al_2O_3}$ far below $T_c$ (inset of Fig. 2f) is also explained by the development of (singlet) superconducting gap and the freeze out of the QP population at a lower $T_{base}$[29,34-38].

Next, using the Nb detector in the same device, we confirm the above interpretation and demonstrate that the conversion efficiency of magnon spin to QP charge can be dramatically enhanced in the vicinity of $T_c$. Figure 3a shows the thermally driven non-local signal $\left[\Delta V_{nl}^{th}\right]^{Nb}$ for the $t_{Nb}$ = 15 nm devices with and without the Al$_2$O$_3$ (spin-blocking) layer at various $T_{base}$ around the superconducting transition of the Nb. In the normal state ($T_{\text{base}}/T_c > 1$), a negative $\left[\Delta V_{nl}^{th}\right]^{Nb}$ (< 0) with several tens of nanovolts is clearly observed. Given $\left[\Delta V_{nl}^{th}\right]^{Pt} > 0$ (see Fig. 2b), this evidences that Nb and Pt have opposite signs in the spin-Hall angle $\theta_{\text{SH}}$, which is in agreement with recent theoretical and experimental studies[38-40]. Intriguingly, upon entering the superconducting state ($T_{\text{base}}/T_c < 1$), a significant enhancement of $\left[\Delta V_{nl}^{th}\right]^{Nb}$ up to *a few microvolts* appears immediately below $T_c$ ($T_{\text{base}}/T_c \approx 0.96$) and then it decays towards zero deep in the superconducting state. We note that there exist visible dips in $\left[\Delta V_{nl}^{th}\right]^{Nb}$ at $\alpha \approx 90°$ and 270° near $T_c$ (Fig. 3a) which are also present for the $Al_2O_3$-inserted control device (Fig. 3b) and thus have nothing to do with the magnon spin-transport and QP iSHE. Similar *spin-independent* signals have been observed in local measurements on NbN/YIG[32] and MoGe/YIG[41] bilayers as well and are explained in terms of Abrikosov-vortex-flow-driven Hall effect under a transverse magnetic field that is close to the upper critical field $\mu_0 H_{c2}$ of (type-II) SC.

To examine the effect of heating power, we measure the $T_{\text{base}}$ dependence of $\left[\Delta V_{nl}^{th}\right]^{Nb}$ (Fig. 3b and 3c) and the normalized $R^{Nb}/R_{T=7\,K}^{Nb}$ (Fig. 3d) at various $I_{\text{dc}}$. As $I_{\text{dc}}$ increases, $T_c$ of the Nb detector is systematically reduced and the transition width becomes broad (Fig. 3d). Accordingly, not only a peak of the $\left[\Delta V_{nl}^{th}\right]^{Nb}$ enhancement shifts to a low $T_{\text{base}}$ but the enhancement regime widens (Fig. 3e). For $I_{\text{dc}} \geq |0.7|$ mA ($J_{\text{dc}} \geq |4.2|$ MA/cm$^2$), the Nb does not turn fully superconducting down to the lowest $T_{\text{base}} = 2$ K studied (inset of Fig. 3d). The corresponding $\left[\Delta V_{nl}^{th}\right]^{Nb}$ then remains non-zero at 2 K and is larger than the normal state value (Fig. 3e). For a quantitative analysis, we plot the normalized voltage $[\Delta V_{nl}^{th}]^{Nb}/[\Delta V_{nl}^{th}]^{Nb}_{T=7K}$ as a function of the normalized temperature $T_{base}/T_c$ in Fig. 3f. We then find that the transition state enhancement of $[\Delta V_{nl}^{th}]^{Nb}/[\Delta V_{nl}^{th}]^{Nb}_{T=7K}$ can reach up to 3 orders of magnitude at the smallest $I_{dc} = |0.1|$ mA ($J_{dc} = |0.6|$ MA/cm$^2$). With increasing $I_{dc}$, its peak amplitude decays rapidly, the full-width-at-half-maximum (FWHM) broadens and the peak is positioned farther away from $T_c$ (inset of Fig. 3f). These results assure that the depressed superconductivity with increasing the heating power has a negative effect on the transition state enhancement of the QP iSHE.

We perform similar measurements on an additional set of the devices with different $t_{Nb}$ (Fig. 4a-4f), comparable to or smaller than the superconducting coherence length $\xi_{SC}$, and thereby strong $t_{Nb}$-dependent superconducting properties (e.g. QP band structure and DOS). Since thin Nb films usually contain a larger amount of grain boundaries, defects, and disorders from the structural inhomogeneity near the growth interface than thick bulk Nb[42,43], the associated scattering effectively weakens electron-electron and electron-phonon interactions, and therefore smearing-out effect of the QP DOS around the gap edge[44]. One would predict a greater enhancement of the QP iSHE if the Nb detector is thicker.

However, experiments give a very different result (Fig. 4g-4i). As $t_{Nb}$ increases, the peak amplitude of $[\Delta V_{nl}^{th}]^{Nb}/[\Delta V_{nl}^{th}]^{Nb}_{T=7K}$ rises until reaching 15 nm and then drops strongly for thicker Nb detectors, leading to a maximum at $t_{Nb} = 15$ nm (Fig. 4j). The width and position of the transition state enhancement, on the other hand, behave as expected for highly and quickly developed coherence peaks in the QP DOS of thick Nb when $T_c$ is crossed: a progressive narrowing of FWHM and a peak shift closer to $T_c$, respectively, with the increase of $t_{Nb}$ (inset of Fig. 4j). The non-trivial $t_{Nb}$-dependent enhancement (Fig. 4j) indicates that there is another key ingredient that controls the enhancement amplitude. That is to say, the exchange spin-splitting field[4-6] which has turned out to considerably modify the QP spin relaxation mechanism via a frozen out of elastic/intravalley spin-flip scattering[4-6]. Below, we discuss how this exchange-field-frozen spin-flip scattering[4-6] is linked to and modifies the QP charge relaxation.

To theoretically describe our results, we first calculate the excited QP spin current density $J_{s0}^{qp}$ at the YIG/superconducting Nb interface as a function of the normalized temperature $T/T_c$ for different values of the magnon spin accumulation $\Delta\mu_m$ relative to the zero-$T$ energy gap $2\Delta_0^{SC}$ (Fig. 5a and 5b). For this calculation, we employ the recent models[29,30] that explicitly take the superconducting coherence factor into account (see Supplementary Section 3 for full details). Note that the characteristic energy of incoherent magnons which excite spin-polarized QPs in the Nb detector is set by $\Delta\mu_m$ and the $T_c$ (or $2\Delta_0^{SC}$) suppression at a larger $\Delta\mu_m$ is inferred from our data set (Fig. 3 and 4). For a quantitative comparison, $J_{s0}^{qp}$ is normalized to its normal state value $J_{s0}$.

The calculated $J_{s0}^{qp}/J_{s0}$ increases remarkably near $T_c$ (0.8$T_c$ – 0.9$T_c$) and it decreases exponentially when $T < 0.8T_c$, reflecting the singularity behaviour in a non-equilibrium population of spin-polarized QPs[29,30,42]. In addition, the peak amplitude of $J_{s0}^{qp}/J_{s0}$ is inversely proportional to $\Delta\mu_m/2\Delta_0^{SC}$ (inset of Fig. 5a and 5b), explaining qualitatively the heating power dependence of the transition-state enhancement (Fig. 3f). Nonetheless, this analysis based on the superconducting coherence factor does not capture the mechanism behind the non-trivial $t_{Nb}$ dependence (Fig. 4j).

We next consider the QP resistivity $\rho_{SC}^{qp}$ (Fig. 5c and 5d) and the volume fraction of QP charge imbalance $v_Q$ (Fig. 5e and 5f) which together determine the effective resistivity $\rho_{SC}^*$ (=$\rho_{SC}^{qp} v_Q$, inset of Fig. 5e and 5f) of the superconducting Nb[38,45]. Here $v_Q = \left(\frac{2\lambda_Q}{w_y}\right) \tanh\left(\frac{w_y}{2\lambda_Q}\right)$[38,45], $w_y$ is the spin-active length of the Nb detector and $\lambda_Q$ is the QP charge-imbalance relaxation length. $\rho_{SC}^{qp}$ and $\rho_{SC}^*$ are normalized by their normal state ones $\rho_0$ and $\rho_0^*$, respectively. We note that if the SC thickness is comparable to or smaller than the QP spin transport length, as relevant to our system[38,46], the QP-mediated iSHE voltage $V_{iSHE}^{qp}$ in the SC can be approximated as $V_{iSHE}^{qp} = \theta_{SH}^{qp} J_{s0}^{qp} \left(\frac{e}{\hbar}\right) \rho_{SC}^* w_y$, where $\theta_{SH}^{qp}$ is the QP spin-Hall angle which is predicted to slightly increase near $T_c$[45,47] (see Supplementary Section 3 for details), $e$ is the electron charge and $\hbar$ is the reduced Planck constant. Consequently, $J_{s0}^{qp}$ and $\rho_{SC}^*$ appear to be governing parameters in $V_{iSHE}^{qp}$.

The most salient aspect of the calculations is that in the vicinity of $T_c$, $v_Q$ dominates the $T$-dependent $\rho_{SC}^*$ over $\rho_{SC}^{qp}$, resulting in $V_{iSHE}^{qp} \propto \left(\frac{2\lambda_Q}{w_y}\right)$ for a given $J_{s0}^{qp}$. This signifies that the QP charge imbalance relaxation is likely responsible for the non-trivial $t_{Nb}$-dependent transition-state enhancement (Fig. 4j) observed in our system.

We thus propose the following mechanism. If QP charge relaxes through the spin-flip scattering $1/\tau_{sf}^{qp}$ and the inelastic scattering $1/\tau_{in}$, and $1/\tau_{sf}^{qp} > 1/\tau_{in}$, the effective relaxation time $\tau_Q^*$ for the QP charge imbalance[48] is given by $\tau_Q^* \approx \frac{4k_B T}{\pi \Delta^{sc}} \sqrt{\frac{\tau_{sf}^{qp} \tau_{in}}{2}}$ with $k_B$ is the Boltzmann constant. Based on the exchange-field-frozen spin-flip scattering[4-6] and its proximity nature[33] in a FMI/SC system, one can reasonably assume $\tau_{sf}^{qp} \propto \Delta E_{ex} \propto \frac{1}{t_{SC}}$. This leads to $\lambda_Q \propto \frac{1}{(t_{SC})^{1/4}}$ and $V_{iSHE}^{qp} \propto \frac{j_{s0}^{qp}}{(t_{SC})^{1/4}}$. Qualitatively, we can understand the $t_{Nb}$-dependent transition-state enhancement (Fig. 4j) in the following manner. When $t_{Nb} \ll \xi_{Nb}$, the superconducting coherence is too weak to excite large QP spin currents across the YIG/Nb interface. In contrast, for $t_{Nb} > \xi_{Nb}$, the exchange spin-splitting-field cannot propagate over the entire depth of such thick Nb and hence the converted QP charge relaxes faster primarily via the spin-flip scattering process. Overall, these two competing effects control the amplitude of the transition-state enhancement by which one would expect a maximum at the intermediate $t_{Nb} \approx \xi_{Nb}$ (around 15 nm for Nb thin films). Note also that the enhancement width and position are determined by $J_{s0}^{qp} \times \rho_{SC}^* \left[\propto \left(\frac{2\lambda_Q}{w_y}\right)\right]$, the latter of which decays rapidly to zero below $T_c$ for a strong superconducting Nb.

To check the validity of this proposal, we experimentally investigate how the transition state enhancement scales with the separation distance $d_s$ between Au/Ru electrical contacts on the exchange-spin-split Nb layer (Fig. 6a, see Supplementary Section 4). Importantly, while the peak position and width of the transition-state enhancement are almost independent of $d_s$ (Fig. 6b and 6c), the peak amplitude increases *exponentially* with the increase of $d_s$ (inset of Fig. 6c), reflecting the characteristics of QP charge-imbalance relaxation effect (see Supplementary Section 4). From the $d_s$-dependent $\left[\Delta V_{nl}^{th}\right]^{Nb}$ (Fig. 6e), we are able to estimate $\lambda_Q$ in the vicinity of $T_c$ ($T/T_c = 0.94 – 0.98$) for the spin-split Nb to be around 90 µm. This is surprisingly a few orders of magnitude larger than either commonly assumed[48] or hitherto reported in Nb films without the presence of spin-splitting fields[49] and thereby should indicate the significantly exchange-field-modified QP relaxation in our system.

Finally, we briefly mention other relevant experiments. It has been previously shown that in all-metallic non-local spin-Hall devices[8], the giant iSHE (~2000 times at most) is created by electrical spin injection from $Ni_8Fe_2$ through Cu into superconducting NbN far below $T_c$ ($T_{base}/T_c = 0.3$) and attributed to the exponentially increasing QP resistivity at a lower $T$. By contrast, a recent experiment has reported that for a YIG/NbN vertical junction[33], the 2–3 times enhanced iSHE voltage by local SSE is measurable only in a limited $T$ range right below $T_c$ ($T_{base}/T_c = 0.96$). In this work, the superconducting coherence factor is pointed out as a main source for such an enhancement and a quantitative description of the data is also provided. In *metallic/conducting* Nb/Ni$_8$Fe$_2$ bilayers[38], a monotonic decay of spin-pumping-induced iSHE appears across $T_c$, indicating no superconducting coherence effect detectable.

## CONCLUSIONS

The new key findings of our study which help understand these puzzling results are as follows. The spin-to-charge conversion mediated by QPs is substantially enhanced in the normal-to-superconducting transition regime, where the interface superconducting gap matches the magnon spin accumulation. The conversion efficiency and characteristics depend crucially on the driving/heating power and the SC thickness, which is understood based on the two competing effects: the superconducting coherence[29,30,42] and the exchange-field-modified QP relaxation[4-6,48]. The validity of these competing mechanisms is experimentally confirmed by spatially resolved measurements with varying the separation of electrical contacts on the spin-split Nb layer. A quantitative reproduction of the result remains an open question for a theory. The coupling between different non-equilibrium imbalances (magnon, spin, charge, heat, magnon-heat, and spin-heat)[4,12] with exchange spin-splitting and the non-linear kinetic equations[4] in the superconducting state should be taken into account rigorously. We speculate that this giant transition-state QP SHE is generic in any FMI/SC system and its efficiency gets even larger especially with two-dimensional (2D) SCs[50] where the exchange spin-splitting can readily proximity-penetrate the entire depth of the 2D SCs. We also anticipate that such a giant spin-to-charge conversion phenomenon (involving non-equilibrium QPs) can be used as an extremely sensitive probe of spin currents in emergent quantum materials[51].

## METHODS

**Device fabrication.** We fabricated the magnon spin-transport devices (Fig. 1b) based on 200-nm-thick single-crystalline YIG films (from Matesy GmbH) by repeating a sequence of optical lithography, deposition, and lift-off steps. Note that these YIG films exhibited a very low Gilbert damping of $0.6 \times 10^{-4}$ at room temperature, determined via ferromagnetic resonance linewidth measurements (by Matesy GmbH, https://www.matesy.de/en/products/materials/yig-single-crystal). We first defined the central Nb detector with a lateral dimension of $9 \times 90\,\mu m^2$, which was grown by accelerated Ar-ion beam sputtering at a working pressure of $1.5 \times 10^{-4}$ mbar. For the control device, a 10-nm-thick $Al_2O_3$ spin-blocking layer was *in-situ* deposited prior to the Nb deposition. We then defined a pair of Pt electrodes of $1.5 \times 50\,\mu m^2$, which were deposited by d.c. magnetron plasma sputtering at an Ar pressure of $4 \times 10^{-3}$ mbar These Pt electrodes are separated by a center-to-center distance $d^{Pt-Pt}$ of 15 μm, which is comparable to the typical $l_{sd}^m$ of single-crystalline YIG films[S1-S3] and also to the estimated values from our Pt-only reference devices with different $d^{Pt-Pt}$ (Supplementary Section 1). The Nb thickness ranges from 10 to 35 nm whereas the Pt thickness is fixed at 10 nm. Finally, we defined Au(80 nm)/Ru(2 nm) electrical leads and bonding pads, which were deposited by the Ar-ion beam sputtering. Before depositing the Au/Ru layers, the Nb and Pt surfaces were gently Ar-ion beam etched for transparent electrical contacts between them.

**Non-local measurement.** We measured the non-local magnon spin-transport (Fig. 1a and 1b) in a Quantum Design Physical Property Measurement System at a temperature varying between 2 and 300 K. A d.c. current $I_{dc}$ in the range of 0.1 to 1 mA was applied to the first Pt using a Keithley 6221 current source and the non-local voltages [$V_{nl}^{Pt}(\alpha)$, $V_{nl}^{Nb}(\alpha)$] across the second Pt and the central Nb are simultaneously recorded as a function of in-plane magnetic-field-angle $\alpha$ by a Keithley 2182A nanovoltmeter. $\alpha$ is defined as the relative angle of $\mu_0 H_{ext}$ (//$M_{YIG}$) to the long axis of two Pt electrodes which are collinear.

## ASSOCIATED CONTENT

**Supporting Information**

The Supporting Information is available free of charge at [TBD].

## AUTHOR INFORMATION

**Author contributions**

K.-R.J. conceived and designed the experiments. The magnon spin-transport devices were fabricated by K.-R.J. with help from J.-C.J., X.Z. and A.M. The non-local transport measurements were carried out by K.-R.J. with help of J.Y. and J.-C.J. K.-R.J. performed the data analysis and model calculation. S.P.P.P. supervised the project. All authors discussed the results and commented on the manuscript, which was written by K.-R.J.

**Competing interests**

The authors declare no competing financial interests.

## ACKNOWLEDGMENTS

This work was supported by the Alexander von Humboldt Foundation.

**FIGURE LEGENDS**

**Figure 1. Non-local magnon spin-transport device with a spin-split superconductor.** (**a**) Schematic illustration of the device layout and measurement configuration. When a d.c. charge current $I_{dc}$ is applied to the right Pt injector, either electrically or thermally driven magnons accumulate in the ferrimagnetic insulator $Y_3Fe_5O_{12}$ (YIG) underneath and diffuse towards the left Pt detector. These magnon ($s = +1$) currents are then absorbed by the left Pt detector, resulting in the electron spin accumulation that is, in turn, converted to a non-local charge voltage $V_{nl}^{Pt}$ via inverse spin-Hall effect (iSHE). Such a conversion process also occurs for the central Nb and thereby $V_{nl}^{Nb}$. However, the conversion efficiency changes dramatically when turning superconducting due to the development of quasiparticle (QP) density-of-states with exchange spin-splitting $\Delta E_{ex}$. Note that in contrast to spin-singlet ($S = 0$) Cooper pairs in coherent ground state, the excited QPs can carry spin angular momentum in the superconducting state. (**b**) Optical micrographs of the fabricated devices with and without a 10-nm-thick $Al_2O_3$ spin-

blocking layer. (**c**) In-plane (IP) magnetization hysteresis $m(H)$ curves of a bare YIG film, measured at a temperature $T$ of $2-300$ K. The inset summarizes the $T$ dependence of the saturation magnetic moment. (**d**) IP magnetic-field-angle $\alpha$ dependence of non-local total voltages $[V_{nl}^{tot}]^{Pt}$ measured with the Pt detector at $I_{dc} = \pm 1.0$ mA at 300 K, for the $t_{Nb} = 15$ nm device. From these, electrically ($[\Delta V_{nl}^{el}]^{Pt}$ in e) and thermally ($[\Delta V_{nl}^{th}]^{Pt}$ in f) driven magnon components are separated (see Main text). Black solid lines in e and f correspond respectively to $\sin^2(\alpha)$ and $\sin(\alpha)$ fits. The estimated magnitude of $[V_{nl}^{el}]^{Pt}$ ($[V_{nl}^{th}]^{Pt}$) is plotted as a function of $|I_{dc}|$ in the inset of e (f), where the black solid line represents a linear fit (quadratic fit). (**g-i**), Data equivalent to d-f but for the control device with the Al$_2$O$_3$ spin-blocking layer.

**Figure 2. Temperature dependence of non-local signals measured by the Pt detector.**
(**a**) Electrically driven non-local voltages $[\Delta V_{nl}^{el}(\alpha)]^{Pt}$, as a function of IP field angle $\alpha$ for the $t_{Nb} = 15$ nm devices with and without the Al$_2$O$_3$ layer, taken at various base temperatures $T_{base}$. The black solid lines are $\sin^2(\alpha)$ fits. (**b**) Data equivalent to A but for thermally driven non-local voltages $[\Delta V_{nl}^{th}(\alpha)]^{Pt}$, along with $\sin(\alpha)$ fits (black solid lines). In these measurements, $I_{dc}$ is fixed at $|0.5|$ mA and the magnetic field $\mu_0 H$ at 5 mT. (**C**) Nb resistance $R^{Nb}$ versus $T_{base}$ plots for the Al$_2$O$_3$-absent and Al$_2$O$_3$-present devices, measured using a four-terminal current-voltage method (using leads 3,4,5,6 in Fig. 1b) while applying $I_{dc} = 0.5$ mA to the Pt injector. A strong suppression of $T_c$ in the absence of the Al$_2$O$_3$ layer (about 1.5 K, at least one order of magnitude larger than expected from stray fields of YIG) indicates the inverse proximity effect[33]; that is, the propagation of YIG-induced exchange spin-splitting into the adjacent Nb. The vertical solid line indicates the superconducting transition temperature $T_c$ of the Nb of the $Al_2O_3$-absent device. Extracted magnitudes of $\left[\Delta V_{nl}^{el}\right]^{Pt}$ (**d**) and $\left[\Delta V_{nl}^{th}\right]^{Pt}$ (**e**) as a function of $T_{base}$ for the $Al_2O_3$-absent and $Al_2O_3$-present devices. In the inset of E, $\Delta\left[\Delta V_{nl}^{th}(T_{base})\right]^{Pt} = \left[\Delta V_{nl}^{th}(T_{base})\right]^{Pt,\ no\ Al_2O_3} - \left[\Delta V_{nl}^{th}(T_{base})\right]^{Pt,\ with\ Al_2O_3}$ is also shown. (**f**) $\left[\Delta V_l^{th}\right]^{Pt,\ no\ Al_2O_3}/\left[\Delta V_{nl}^{th}\right]^{Pt,\ with\ Al_2O_3}$ as a function of $T_{base}$ and $T_{base}/T_c$ (inset).

**Figure 3. Giant enhancement of non-local signals in the transition state of the Nb detector.** (**a**) Thermally driven non-local voltages $\left[\Delta V_{nl}^{th}(\alpha)\right]^{Nb}$ as a function of IP field angle $\alpha$ for the $t_{Nb} = 15$ nm devices with and without the $Al_2O_3$ layer, taken at $I_{dc} = |0.5|$ mA around $T_c$ of the Nb. The black solid lines are $\sin(\alpha)$ fits. (**b** and **c**) Data equivalent to A but at $I_{dc} = |0.10|$ mA (b) and $I_{dc} = |0.60|$ mA (c), respectively, for the $Al_2O_3$-absent device. (**d**) Normalized Nb resistance $R^{Nb}/R_{T=7K}^{Nb}$ versus $T_{base}$ plots for the $Al_2O_3$-absent device, measured using a four-terminal current-voltage method (using leads 3,4,5,6 in Fig. 1b) with varying $I_{dc}$ in the Pt injector. The critical temperature $T_c$ is defined as the point where $R^{Nb} = 0.5 R_{T=7K}^{Nb}$. The inset summarizes the measured $T_c$ as a function of $I_{dc}$ (or $J_{dc}$). (**e**) Estimated magnitude of $\left[\Delta V_{nl}^{th}\right]^{Nb}$ as a function of $T_{base}$ for the $Al_2O_3$-absent device. (**f**) $\left[\Delta V_{nl}^{th}\right]^{Nb}/\left[\Delta V_{nl}^{th}\right]_{T=7K}^{Nb}$ versus $T_{base}/T_c$ plot. The inset displays the $|I_{dc}|$ (or $|J_{dc}|$) dependence of the peak amplitude, width, and position.

**Figure 4. Nb thickness dependence of the giant transition-state enhancement.**

Representative non-local signals $\left[\Delta V_{nl}^{th}(\alpha)\right]^{Nb}$ as a function of IP field angle $\alpha$ for the Al$_2$O$_3$-absent devices with different $t_{Nb}$ of 10 (**a** and **b**), 20 (**c** and **d**), and 35 nm (**e** and **f**), taken above (yellow background) and *immediately* below (blue background) $T_c$ of the Nb layer. $\left[\Delta V_{nl}^{th}\right]^{Nb}/\left[\Delta V_{nl}^{th}\right]^{Nb}_{T=7K}$ versus $T_{base}/T_c$ plots for $t_{Nb}$ = 10 nm (**g**), $t_{Nb}$ = 20 nm (**h**), and $t_{Nb}$ = 35 nm (**i**). In the insets of (**g-h**), the associated $R^{Nb}/R^{Nb}_{T=7K}$ and $\left[\Delta V_{nl}^{th}\right]^{Nb}$ are plotted as a function of $T_{base}$. (**j**) $t_{Nb}$-dependent peak amplitude, width (inset), and position (inset).

**Figure 5. Theoretical identification of origins for the giant transition-state enhancement.** (**a** and **b**) Normalized QP spin current density $J_{s0}^{qp}/J_{s0}$ at the YIG/superconducting Nb interface as a function of the normalized temperature $T/T_c$. In this calculation, we use various values of the magnon spin accumulation $\Delta\mu_m$ relative to the zero-$T$ energy gap $2\Delta_0^{SC}$. Note that $\Delta\mu_m$ and $2\Delta_0^{SC}$ are both inferred from our data set (Fig. 3 and 4) using relevant theories (see Supplementary Section 3). Each inset summarizes the peak amplitude of $J_{s0}^{qp}/J_{s0}$ versus $\Delta\mu_m/2\Delta_0^{SC}$. (**c** and **d**) Normalized QP resistivity $\rho_{SC}^{qp}/\rho_0$ as a function of $T/T_c$. (**e** and **f**) Volume fraction of QP charge imbalance $v_Q$ as a function of $T/T_c$. In this calculation, we use three different QP charge-imbalance relaxation lengths $\lambda_Q$ = 15, 150, and 1500 nm in the low-$T$ limit ($T/T_c \ll 1$). Insets of (e and f) display the normalized effective resistivity $\rho_{SC}^*/\rho_0^*$ (=$\rho_{SC}^{qp}v_Q/\rho_0$) of the superconducting Nb.

**Figure 6. Spatial profiling of the giant transition-state enhancement.** (**a**) Optical micrographs of the fabricated devices, in which only the separation distance $d_s$ of Au/Ru

electrical contacts on the 15-nm-thick Nb layer varies from 10 to 60 µm. (**b**) Thermally driven non-local signals $\left[\Delta V_{nl}^{th}\right]^{Nb}$ as a function of the base temperature $T_{\text{base}}$ for the devices with different $d_s$. In these measurements, $I_{\text{dc}}$ is fixed at |0.5| mA and the magnetic field $\mu_0 H$ at 5 mT. The inset shows the normalized Nb resistance $R^{Nb}/R_{T=7K}^{Nb}$ versus $T_{\text{base}}$ plot, which confirms nearly identical superconducting transition $T_c$ of the Nb layer. Note that a relatively higher $T_c$ of the 15-nm-thick Nb layer in these devices than that of the prior device (Fig. 3d) is due to the better initial base pressure ($< 1 \times 10^{-9}$ mbar) before film deposition. (**c**) $\left[\Delta V_{nl}^{th}\right]^{Nb}/\left[\Delta V_{nl}^{th}\right]_{T=7\,K}^{Nb}$ as a function of $T_{\text{base}}/T_c$. The right inset displays the $d_s$ dependence of the peak amplitude, width, and position: the black solid line is an exponential fit whereas the black dashed lines are given by guides to the eye. A magnified plot of the peaks is also shown in the left inset. $d_s$-dependent $\left[\Delta V_{nl}^{th}\right]^{Nb}$ above (**d**) and *immediately* below (**e**) $T_c$ of the Nb layer. In E, the solid lines are fitting curves to estimate the QP charge-imbalance relaxation length $\lambda_Q$ (see Supplementary Section 4).

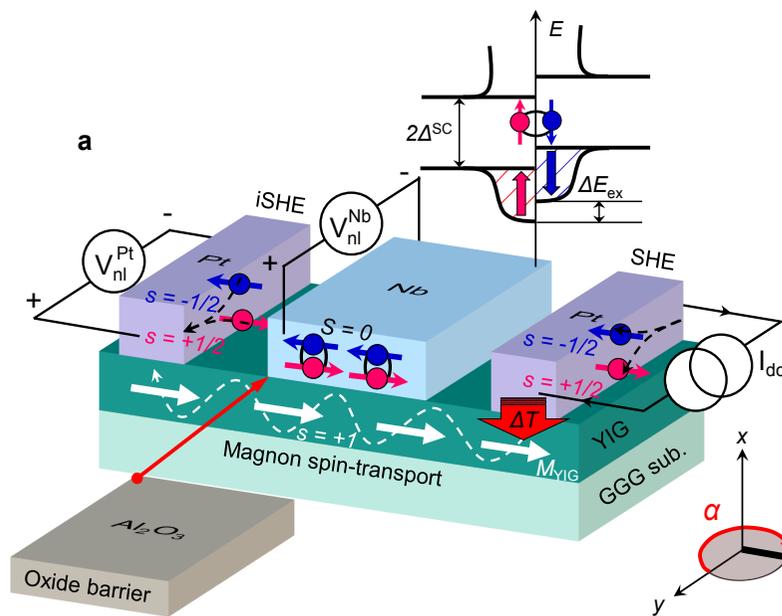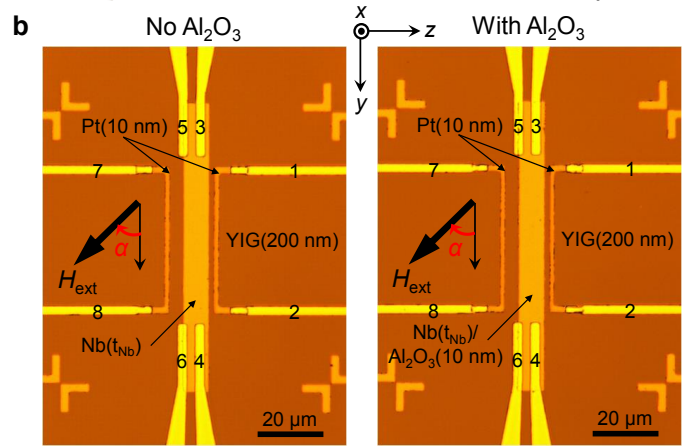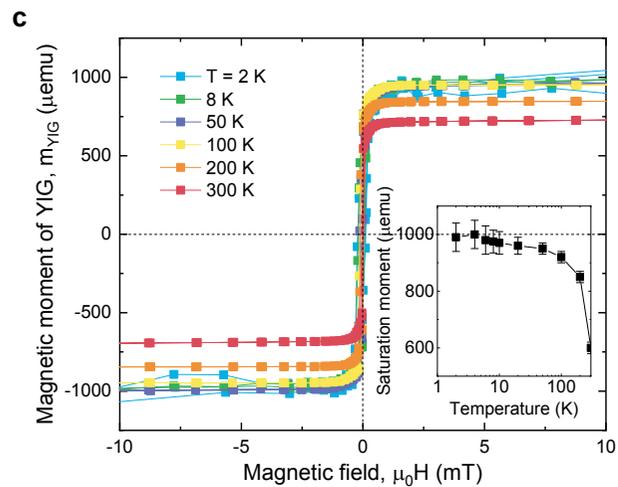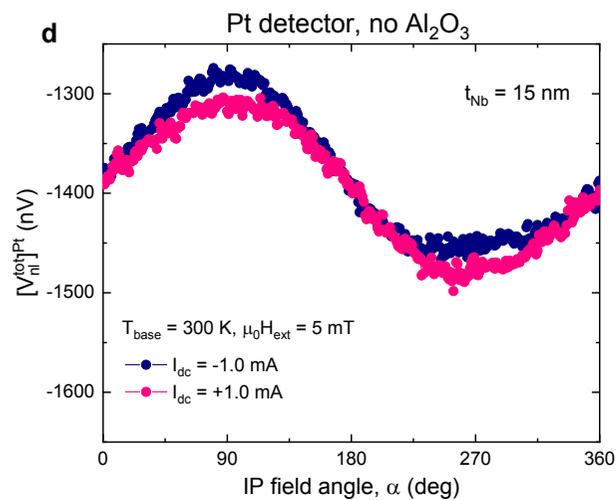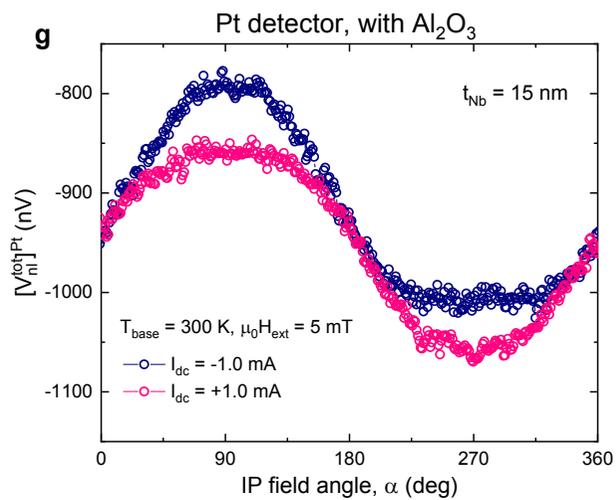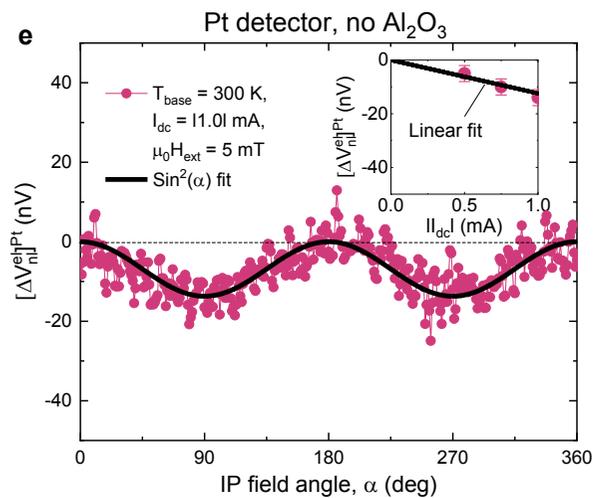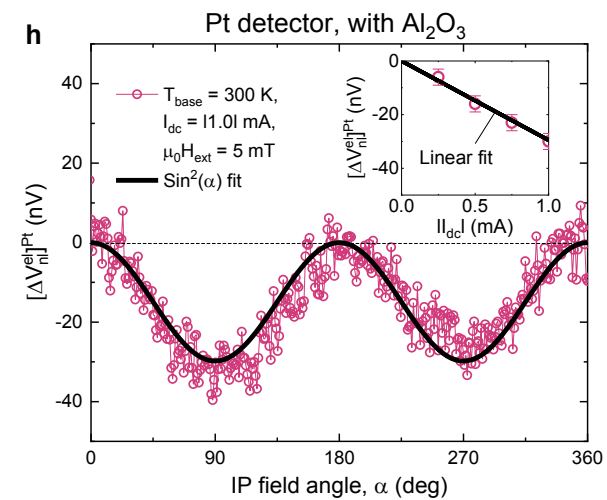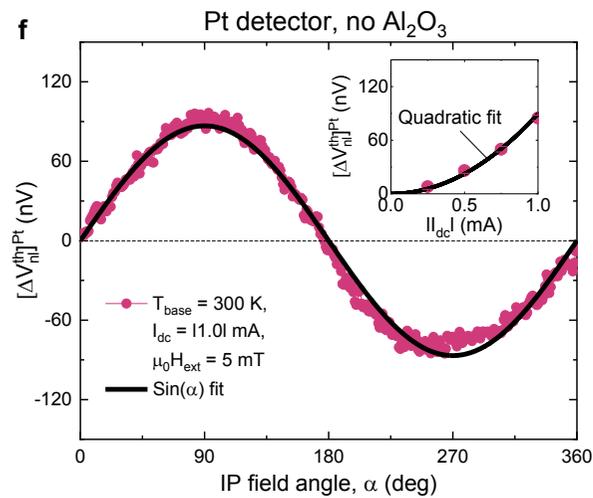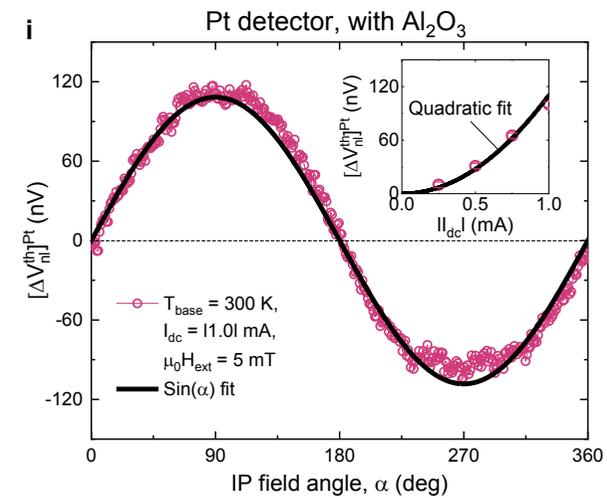

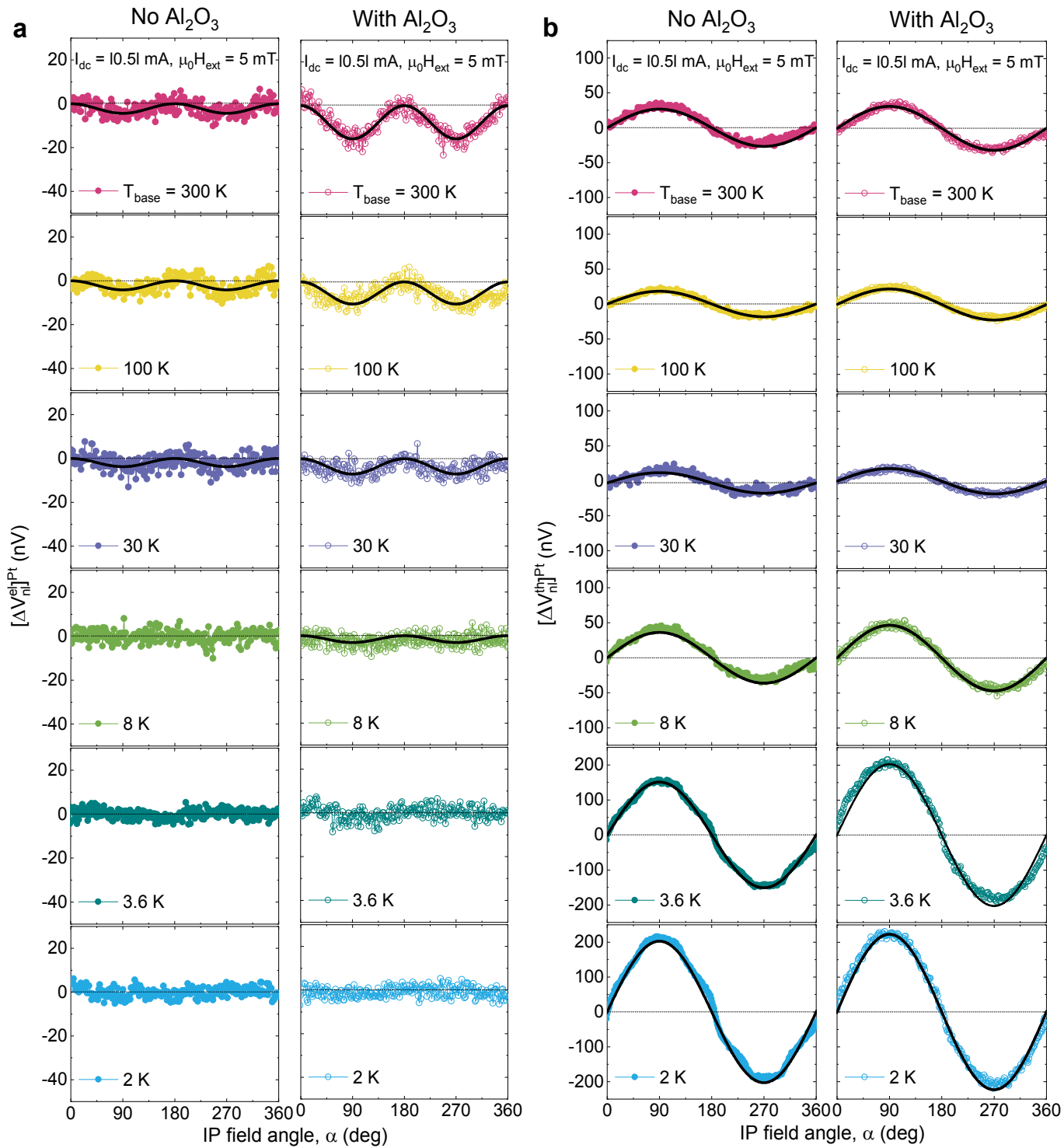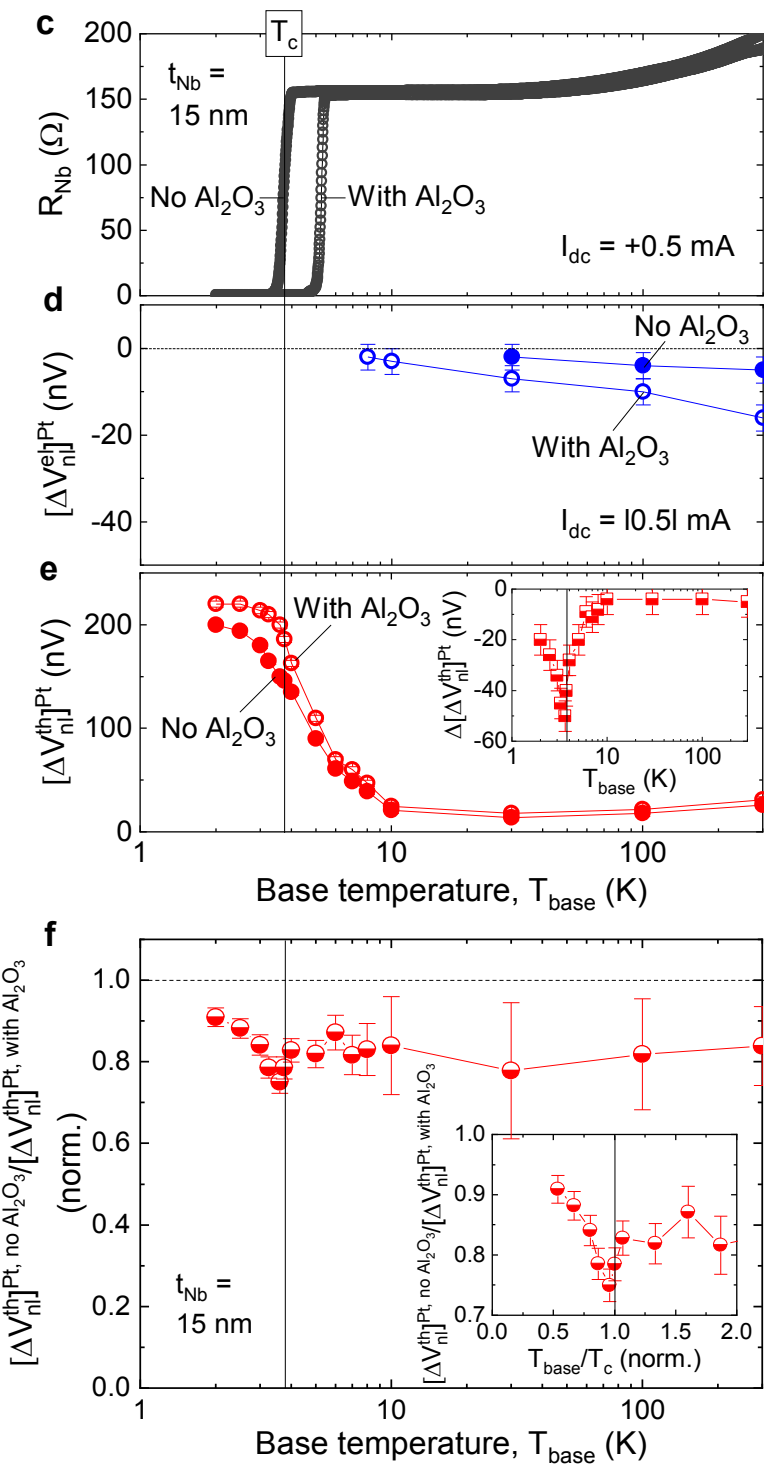

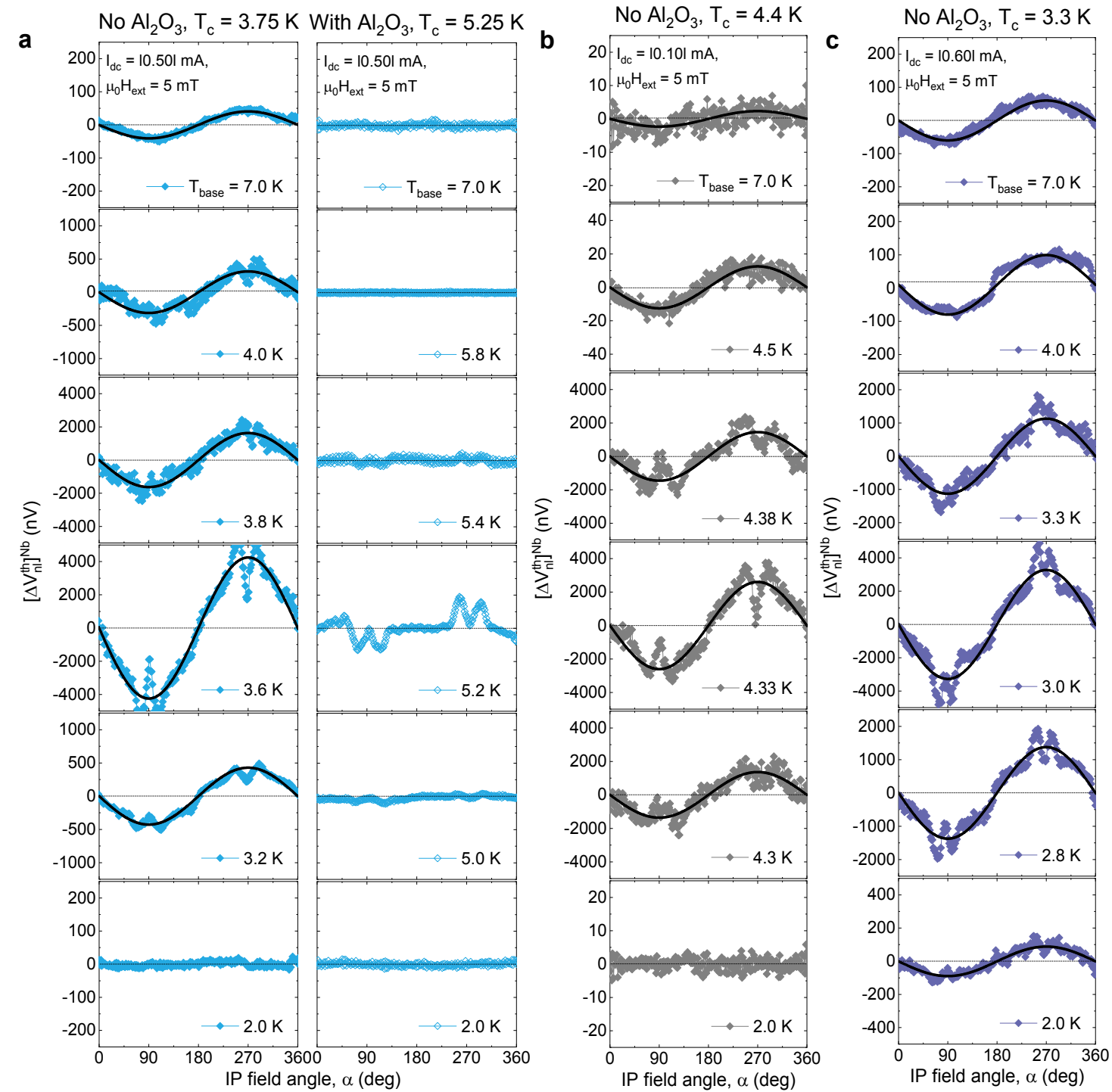

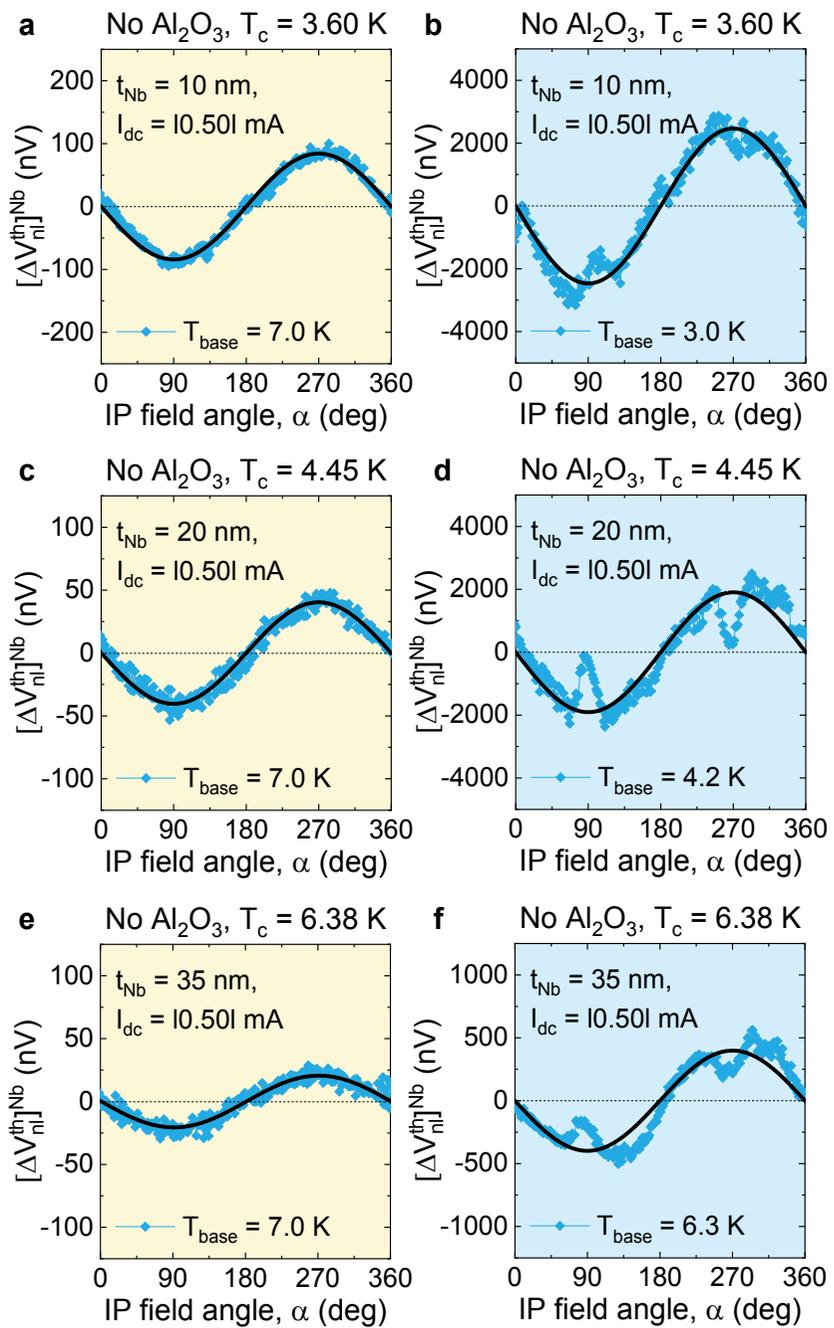
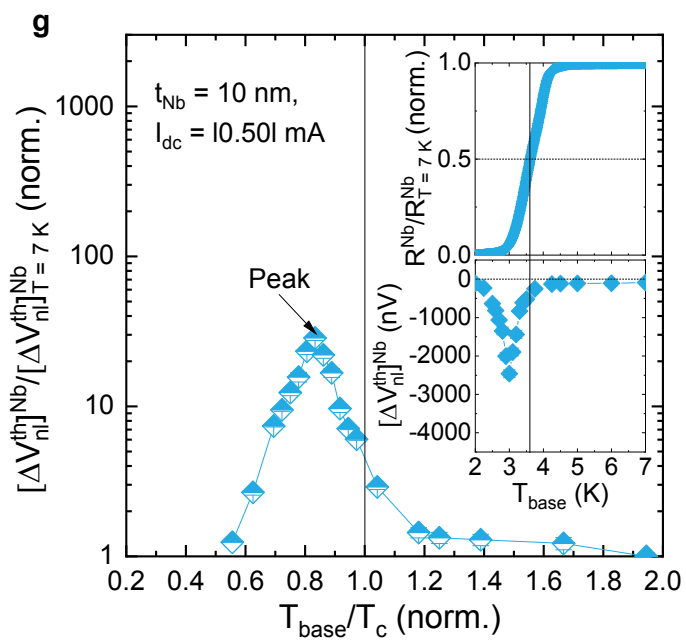
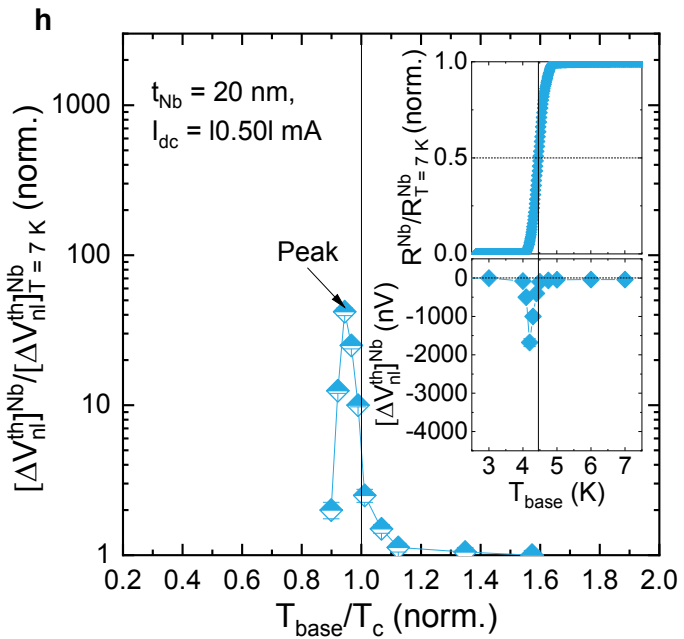
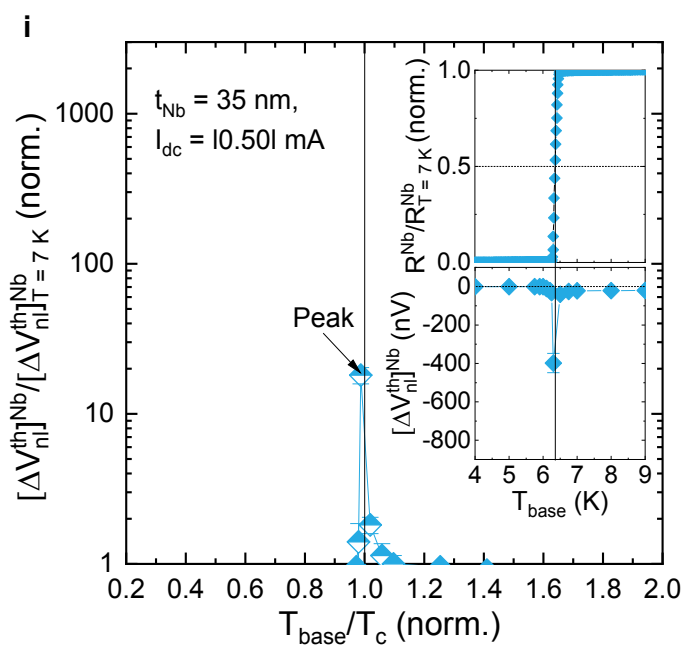
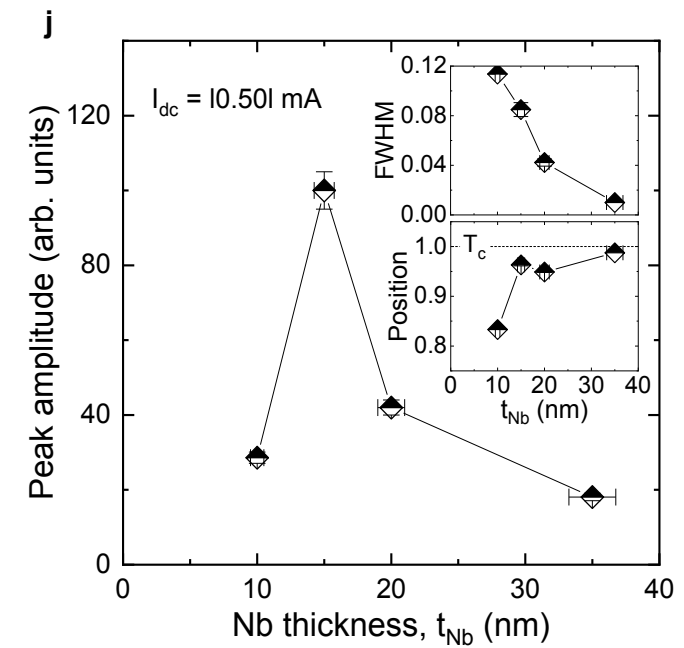

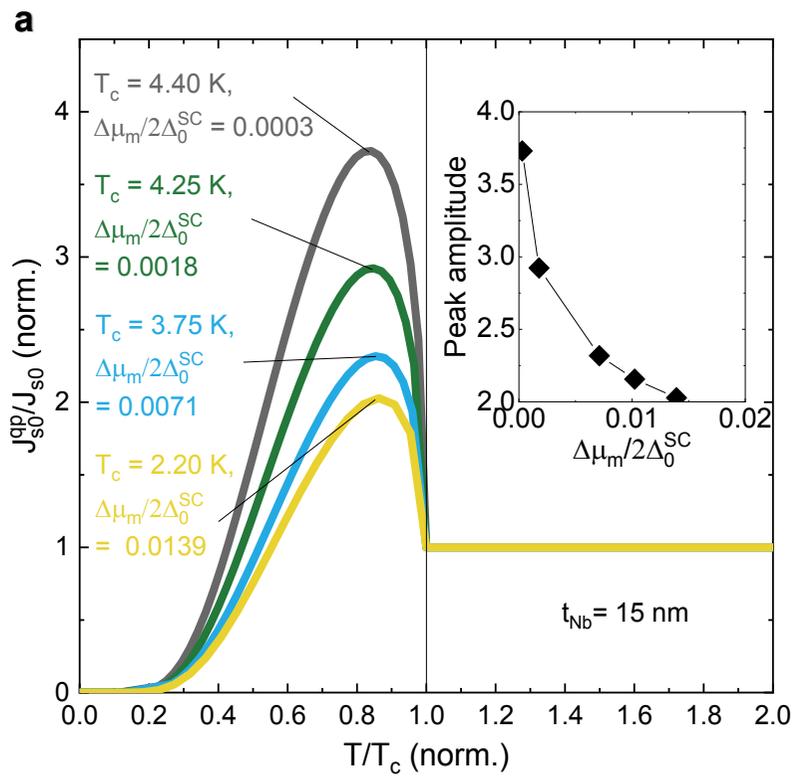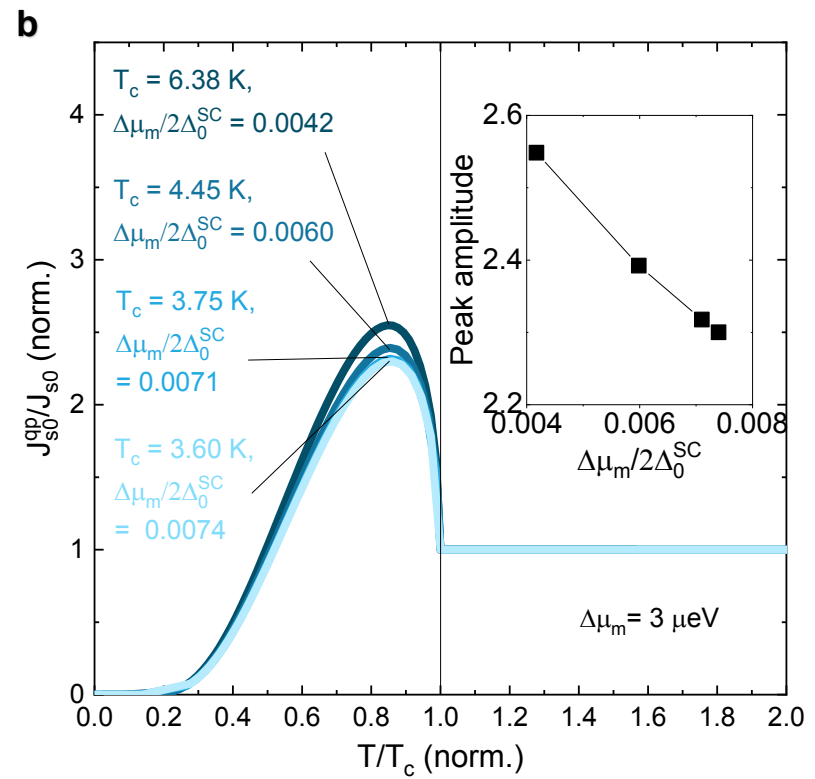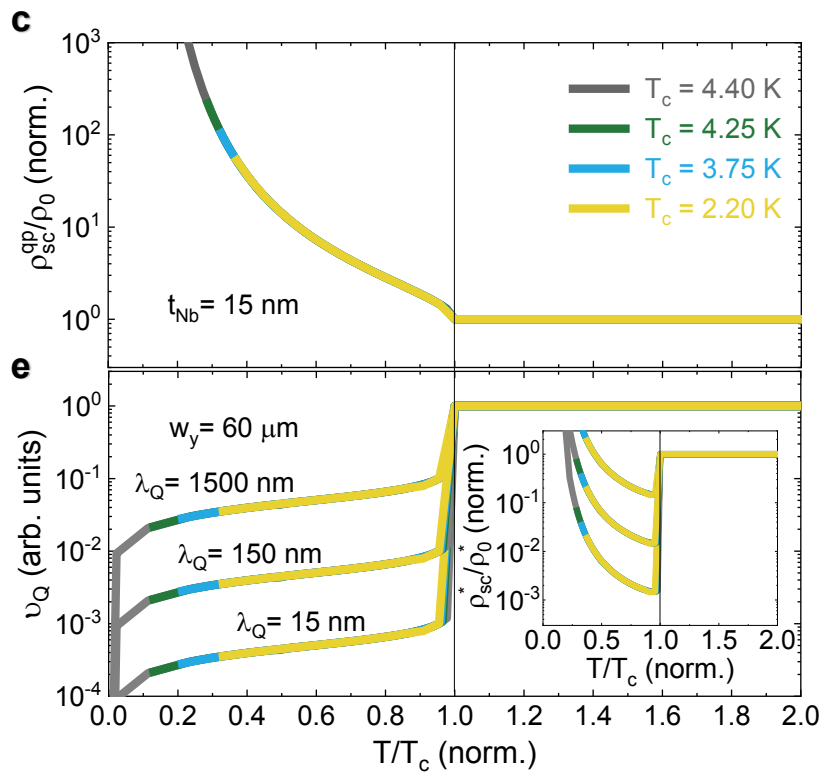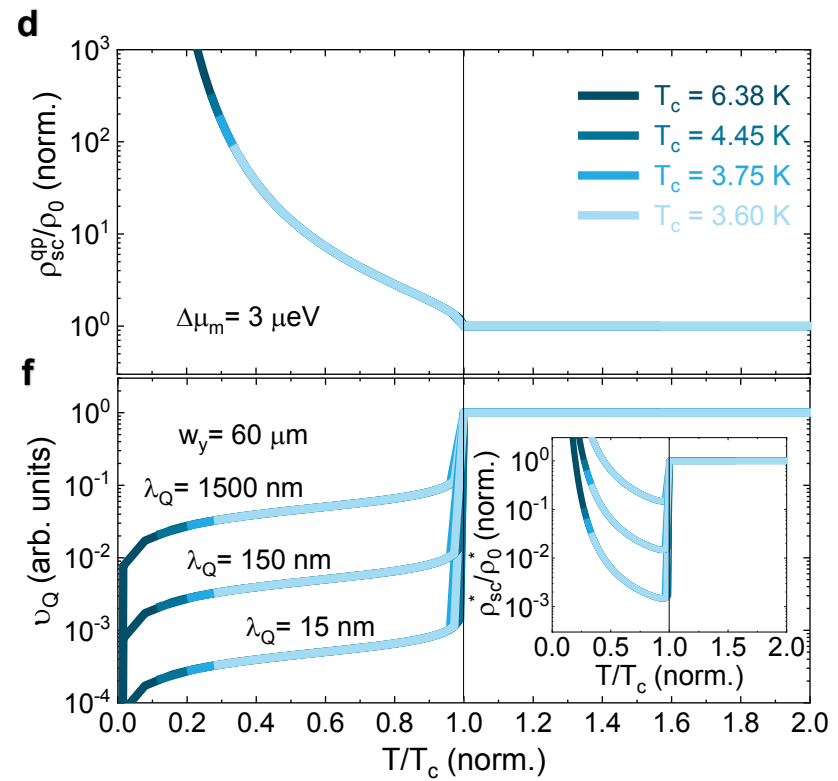

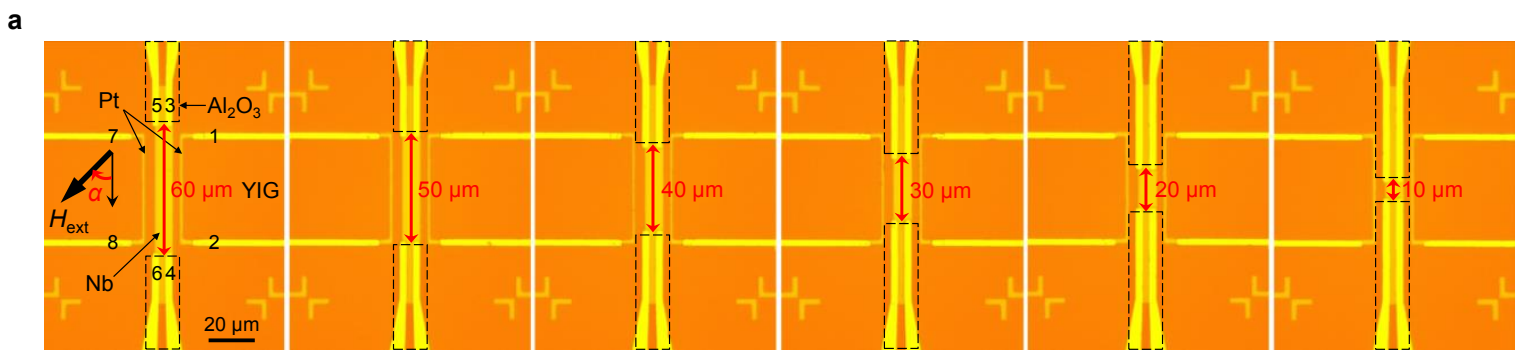
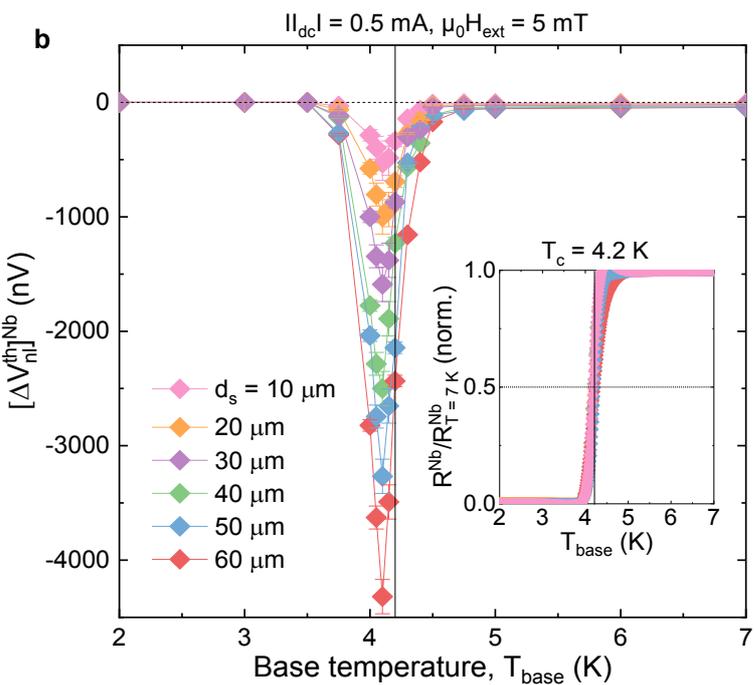
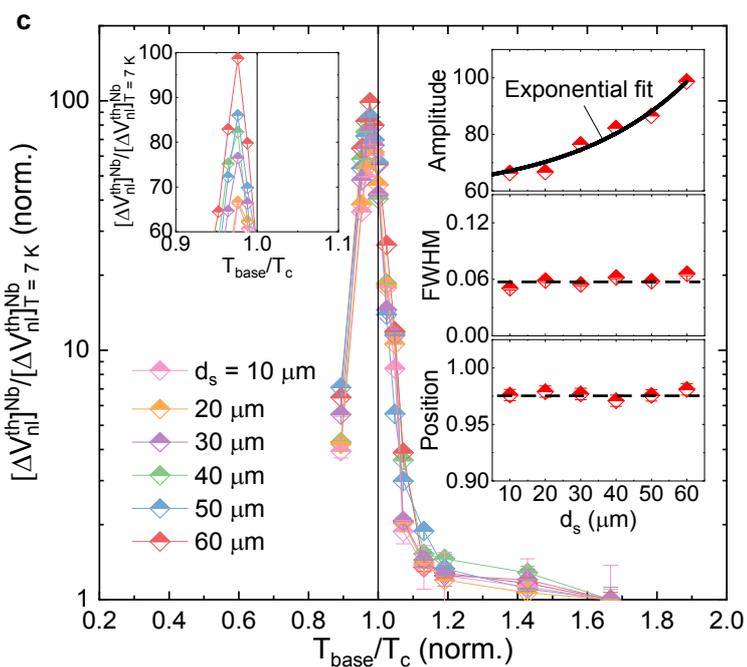
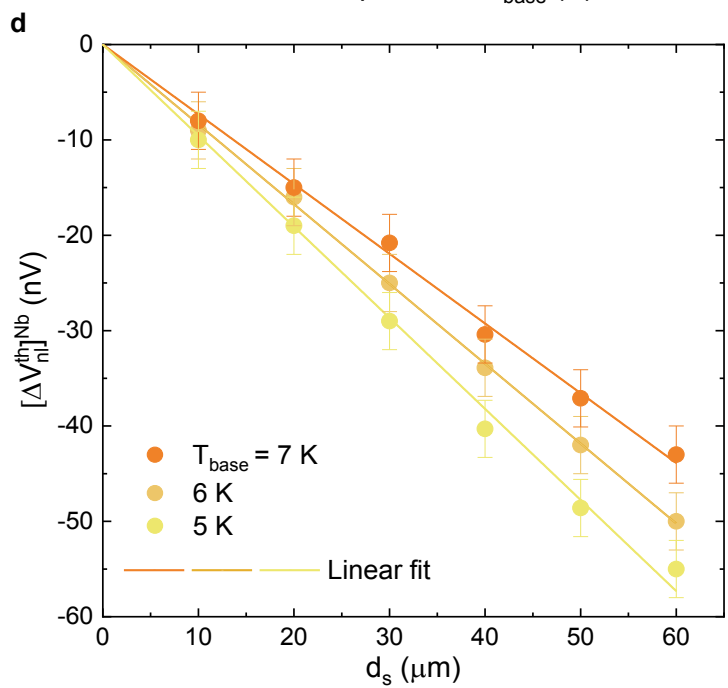
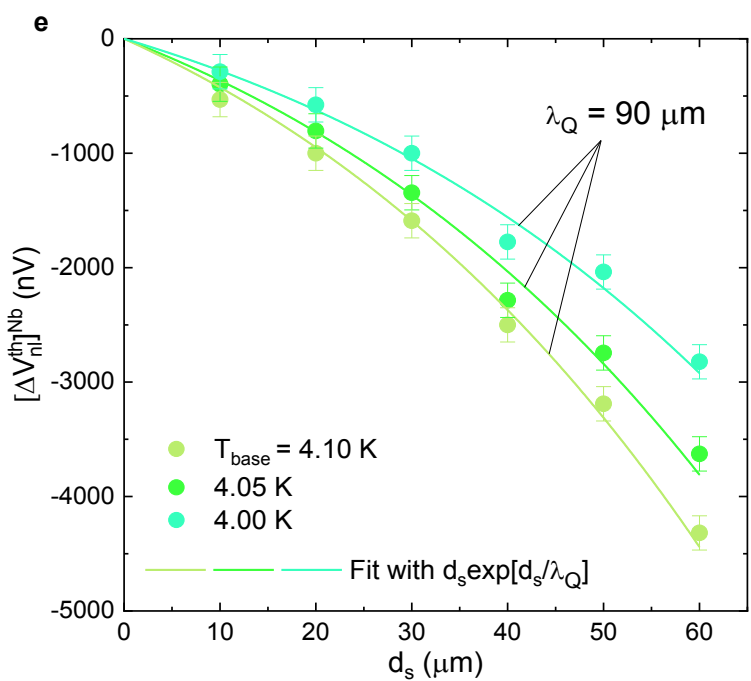